\documentclass[10pt]{article}
\usepackage{amsmath,amssymb,amsthm}
\usepackage{graphicx}
\usepackage{cite}
\title{ Casimir Energy between a Sinusoidally Corrugated Sphere and a Plate Using Proximity Force Approximation}
\author{M. R. Setare, A. Seyedzahedi \\
\small{\emph{Department of Science, University of Kurdistan, Sanandaj, Iran.}}\\
\small{Email address: \emph{rezakord@ipm.ir}} }

\date{}

\theoremstyle{plain}

\begin{document}

\maketitle

\setlength{\parindent}{0pt}

\begin{abstract}
\noindent
In this paper aiming to obtain the Casimir energy between a sinusoidally corrugated sphere and a plate, we first present a derivation of the sphere-plate Casimir force obtained by applying proximity force approximation (PFA). One may investigate this problem by using the scattering matrix method. Despite that this method achieves the exact result, but difficulties of applying this formalism make the procedure complicated. We use the PFA to achieve the purpose in a truly explicit manner. Considering the sphere to have different kinds of sinusoidally corrugations, we obtain the effect of deformation on the Casimir interaction energy. We suppose that $ a/r \leqslant 0.00755$ for the validity of PFA and an experimental accuracy goal of $1\verb"%"$ in the case of sphere with radius $r$ at a minimal distance $a$ from a plate \cite{0601094}. The effect of finite conductivity would be non negligible for such short distances. To reduce the inaccuracy of the result obtained by PFA, we have also investigate thermal corrections of the Casimir force.

\emph{keyword:}  Casimir energy, Proximity force approximation, Corrugated sphere.

\end{abstract}

\section{Introduction}
The dominant interaction between two neutral plates with micron or sub-micron range separation distance is the Casimir effect \cite{casimir}. This tremendous effect with macroscopic manifestation has got a significant role in the miniaturized physical systems. This quantum effect originates from modification of the vacuum fluctuations. It have been investigated for some different kinds of geometries, number of spatial dimensions, fields and boundary conditions (see \cite{Milton,Advances} for a review). Theoretical studies and lots of experiments have been performed to verify the realization of the Casimir effect. Considering that maintaining parallel plates has been a challenge experimentally, most of the experiments have been done by using the configuration of sphere-plate instead of two parallel plates geometry \cite{Lamoreaux,Mohideen,Harris}.
According that sphere-plate geometry is a highly relevant experimental configuration, it has got very fast theoretical and experimental advancement.
% we choose the PFA to get a sense for the physics.

% Some of the studies containing the calculations of the Casimir energy for the electromagnetic field in different geometries

Some of the studies containing the sphere-plate electromagnetic vacuum energy has been extremely done since 2007 \cite{Emig3, Emig4, Neto1,Canaguier}.
These groups have performed exact Casimir calculations for sphere-plate geometries that were based on the scattering matrix methods. In these investigations the plate is supposed to be perfectly conducting whereas on the sphere the electromagnetic field satisfied either perfect conductor boundary conditions or a dielectric with constant. Semitransparent boundary conditions and validity of the proximity force approximation has been investigated in \cite{ Milton1}.
The repulsion in the sphere-plate geometry has been studied with the aim of finding the constraints on the repulsion caused by magnetic permeability of the sphere has been studied in \cite{ Pirozhenko3} by using the approaches of the papers \cite{ Emig3,Emig4, Neto1,Canaguier}. Even Non-equilibrium Casimir force for the sphere-plate geometry has been treated both analytically and numerically in \cite{ Matthias}. Calculations of the Casimir force between a plate and a nanostructured surface at finite temperature in the framework of the scattering theory has been presented in \cite{1212.5479}. The Casimir-Polder forces between an atom and a surface with arbitrary uniaxial corrugations has been presented applying a technique that is fully nonperturbative in the height profile of the corrugation \cite{0810.3480}. And by using height distribution function, effect of roughness or surface modulations on the distance dependence of power-law interactions between curved objects at proximity has been investigated recently in\cite{1303.2499}.
%Some of the mentioned investigations have been done by applying path integral approach. These analysis lead in the exact result, but there are some difficulties in applying this formalism.
Applying worldline numerics \cite{LangfeldLangfeld,Moyaerts}, the Casimir interaction energies for the sphere-plate and cylinder-plate configuration due to the scalar-field fluctuations under the Dirichlet boundary conditions for a wide range of curvature parameters has been examined in \cite{0601094}. Based on a high-precision calculation using worldline numerics, Gies and Klingm$\ddot{u}$ller have determined the validity bounds of the proximity force approximation quantitatively. They observed that for the accuracy goal of $1\verb"%"$, the PFA is valid for $a/r\leqslant 0.00755$. Where $a/r$ is called the curvature parameter, $r$ is radius of the sphere at a minimal distance $a$ from a plate. Analysis shows that, using PFA to obtain plate-sphere result from the plate-plate conclusion for corrugations with wave length $\lambda$ is a good approximation until $r a \gg \lambda^2$ and the amplitude of corrugation is smaller than the other length scales (see \cite{0603120} and references there in). Applying the proximity force approximation for corrugated surfaces may not lead to precise expressions obviously. But using this approximation conducts in a simple procedure to achieve the purpose and emit difficulties appear in the exact investigations.
% We study the lateral dependence of the Casimir energy for different corrugated gratings of arbitrary periodic profile.
% It is worth mentioning that Casimir force strongly depends on the shape and orientation and any deformation in the interacting surfaces may lead in a change in this interaction.
\par
Since sphere-plate configuration is experimentally very important and the Casimir force strongly depends on the interaction bodies' properties, state of their surface, their shape and many other factors, this paper is devoted to a calculation of the Casimir force between a sinusoidally corrugated sphere and a plate. In order to achieve this purpose, we do not develop new method of calculation here. Organization of the paper is as follows:
with respect to the regularized zero point energy density for two parallel plates, we present a derivation for the sphere-plate geometry in sec. 2. One may obtain the Casimir energy using the scattering matrix methods. This method concludes in the exact result but there are several difficulties during the procedure. Therefore we use the proximity force approximation. As we mentioned this approximation is applicable for large sphere with curvature parameters smaller than $0.00755$ for the accuracy goal of $1\verb"%"$. For such short distances the effect of finite conductivity would be non negligible. With respect to the finite conductivity we obtain the corresponding corrections in the Casimir interaction energy. In sec. 3. we suppose the sphere to be corrugated and with the aid of additive summation of the Casimir energy we consider the effect of deformation on the Casimir effect. During this section, we introduce azimuthal and polar sinusoidally corrugated spheres and finally we focus on a golf ball like corrugation and obtaining it's associate energy. We also consider the thermal corrections of the Casimir force to reduce the inaccuracy of the result obtained by PFA in sec. 4. The investigated configurations, considering roughness of the spheres' surfaces and Finite temperature analysis, may be useful in designing nanoscale devices and microelectromechanical systems.

\par
\label{}

\section{The Casimir Interaction Energy for Sphere-Plate Geometry}
The approximate Casimir energy between two curved surfaces placed at a short separation distance can be obtain as a sum of energies between a pair of small parallel plates \cite{Blocki,Klimch}. Considering this approximation we try to use two parallel plates' Casimir energy to obtain the Casimir force for the sphere-plate geometry.
As it is well known The electromagnetic Casimir energy density for two parallel perfect conductor plates with separation distance $H$ is given by \cite{casimir,Milonni,Bordag}
\begin{eqnarray}\label{E}
\mathfrak{E}_{\mathrm{pp}}(H)=- \frac{\pi^{2} }{720 H^{3}},
\end{eqnarray}
and consequently the normal Casimir force density for this configuration is
\begin{eqnarray}\label{F}
\mathfrak{F}_{\mathrm{pp}}(H)=- \frac{\pi^{2}}{240 H^{4}}.
\end{eqnarray}
Let us consider a sphere with radius $r$ above a plate, located in the shortest distance $a$. The Casimir force between the sphere and plate can be obtained by integrating the Casimir force density over the half-sphere which is opposite to the plane
\begin{eqnarray} \label{force}
\hspace{-1 cm}F_{\mathrm{sp}}=\int  \mathfrak{F}_{\mathrm{pp}}(H) ds,
\end{eqnarray}
where the subscript sp stands for the sphere-plate geometry. (Similar investigation has been done for cylinder-plate geometry in \cite{Teo}.)
One can use polar coordinates to determine the normal distance from a point of the sphere with parameter $\theta$  to the plate as $H =a + r (1-cos\theta)$.
Thus the Casimir force for this geometry can be written as
\begin{eqnarray} \label{normal}
\hspace{-1 cm}F_{\mathrm{sp}}=-\frac{\pi^3 r^2}{120}\int_0 ^{\pi/2} \frac{sin\theta}{[a+r(1-cos\theta)]^4} d\theta,
\end{eqnarray}
considering that $a<<r$, the Casimir interaction force reads
\begin{eqnarray} \label{normal2}
\hspace{-1 cm}F_{\mathrm{sp}}=-\frac{\pi^3 r}{360 a^3}.
\end{eqnarray}
Integrating the normal Casimir force with respect to the separation distance $a$, the Casimir interaction energy between a sphere and a plate without corrugations is given by
\begin{eqnarray} \label{E-sp}
\hspace{-1 cm}E_{\mathrm{sp}}=\int_{a}^\infty F_{\mathrm{sp}}  da =
-\frac{\pi^3  r}{720 a^2}.
\end{eqnarray}
It is worth mentioning that Lambrecht and Marachevsky have obtained the exact Casimir force between two arbitrary periodic dielectric gratings. They have found that considering finite conductivity gives a smaller deviation of the exact force from the PFA prediction than the calculation for perfect mirrors \cite{Astrid2}. Considering finite conductivity of metals, corresponding corrections may be included in the mentioned Casimir energy per unit area for two parallel perfect conductor plates \cite{Dzya,Schwinger1,Bezerra1}
\begin{eqnarray} \label{delta}
\mathfrak{E}_{\mathrm{pp}}(H)\thicksim - \frac{\pi^{2}}{720 H^{3}} \bigg(1+ \sum_{n=1}^{4} C_n \big(\frac{\lambda_p}{2 \pi H}\big)^n\bigg),
\end{eqnarray}
where $\lambda_p$ is the plasma wavelength and the coefficients $C_n$ are as follows
\begin{eqnarray} \label{C}
C_1 = -4, C_2 = \frac{72}{5}, C_3=-\frac{320}{7}(1-\frac{\pi^2}{210}), C_4=\frac{400}{3}(1-\frac{163 \pi^2}{7350}). \nonumber
\end{eqnarray}
Eq. (\ref{delta}) is applicableat the separations $H \geq \lambda_p$. Imposing PFA approximation on the correction terms leads in the
following correction in the force derived by Eq. (\ref{normal2})
\begin{eqnarray}\label{delta F}
\Delta F_{\mathrm{sp}}\thicksim - \frac{\pi^{3} r}{120} \sum_{n=1}^{4}\,C_n \,(\frac{\lambda_p}{2 \pi})^{n}\frac{1}{ a^{n+3}},
\end{eqnarray}
integrating the correction of the Casimir force with respect to the separation distance, corresponding correction to the Casimir energy due to the finite conductivity can be obtained
\begin{eqnarray} \label{conductivity}
\hspace{-1 cm} \Delta E_{\mathrm{sp}}\thicksim -\frac{\pi^{3} r}{120}
\sum_{n=1}^{4}\,C_n \,(\frac{\lambda_p}{2 \pi})^{n}\frac{1}{(n+2) a^ {n+2}}.
\end{eqnarray}

\section{Imposing Effect of Deformation on the Sphere-Plate Casimir Interaction Energy}
The Casimir interaction strongly depends on the shape and orientation, therefore any kind of deformation in the interacting surfaces may lead in a modification in the this interaction. We want to develop the PFA for a sphere with periodic corrugations above a smooth plate.
This description is valid only to study deformations with large wavelengths. Hence the roughness effect is obtain to be a small correction to the Casimir energy \cite{Genet,Maia1,Maia2}.
\par
\begin{figure}[h]
\hspace{+2 cm}\includegraphics[width=0.65 \columnwidth]{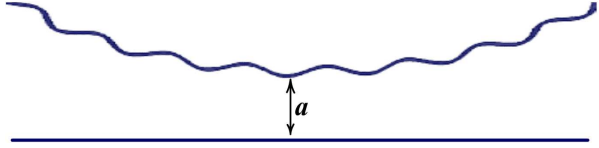}
%\vspace{0.5in}
\caption{A cross section of azimuthal corrugated sphere above the plate with $a<<r$ for applicability of PFA. Where $r$ is radius of the sphere and $a$ is the shortest separation distance.}
\label{sphere-plate}
\end{figure}
\subsection{Effect of Azimuthal Corrugation}
At this point we concentrate on the the sinusoidally corrugations. As the first example assume that the sphere is corrugated in the azimuthal direction (see Fig. (\ref{sphere-plate})). The following function describes the radius of the corrugated sphere
\begin{eqnarray} \label{corrugation1}
r'= r + A \sin(\mu \varphi),
\end{eqnarray}
where $r'$ and $r$ are radiuses of sphere with and without corrugation respectively, $A$ is the corrugation amplitude and $\mu$ is the corrugation's frequency, it is necessary to be a positive integer. Considering the additive summation of the Casimir energy of the sphere-plate geometry without corrugation obtained in the Eq. (\ref{E-sp}) and knowing that according to Eq. (\ref{corrugation1}) the shortest distance between corrugated sphere and plate is still $a$, the Casimir energy For corrugated sphere (with large corrugation wavelength) above the smooth plate can be obtained
\begin{eqnarray} \label{E-corrugated1}
& \hspace{-4 cm} E_{\mathrm{sp}}^{\mathrm{cor}} =\frac{1}{2\pi}\int_{0}^{2\pi} E_{\mathrm{sp}}(\varphi) d\varphi \nonumber \\
&= -\frac{\pi^3 }{720 a^2}\frac{1}{2 \pi} \int_{0}^{2\pi} ( r + A \sin(\mu \varphi)) d\varphi\nonumber \\
&= -\frac{\pi^2 }{720 a^2} \big[\pi r \mu + A \sin^{2}(\mu \pi) \big]/ \mu.
\end{eqnarray}
In which, we have used the face that all the radiuses correspond to different azimuthal angels introduced by Eq. (\ref{corrugation1}) have equal probability  \cite{Chen,Bordag,Emig,Klimchitskaya}.
As mentioned before $\mu$ is an integer and therefore the effect of azimuthal corrugation will be canceled in the  Eq. (\ref{E-corrugated1}). Indeed the Casimir energy takes the form of sphere-plate energy without corrugation ( Eq. (\ref{E-sp})).
\\
Correction of finite conductivity Eq. (\ref{conductivity}) in this case of corrugation yields in the
\begin{eqnarray} \label{azimuthal}
\hspace{-1 cm} \Delta E_{\mathrm{sp}}^{\mathrm{cor}}\thicksim -\frac{\pi^{3}}{120}(r+\frac{A \sin^2(\mu \pi)}{\mu \pi})
\sum_{n=1}^{4}\,C_n \,(\frac{\lambda_p}{2 \pi})^{n}\frac{1}{(n+2) a^ {n+2}} ,
\end{eqnarray}
considering that $\mu$ is integer Eq. (\ref{azimuthal}) returns to Eq. (\ref{conductivity}) and the effect of azimuthal corrugation in the finite conductivity correction will be emitted.

\par
\subsection{Effect of polar Corrugation}
For another simple example of sinusoidal corrugations, we describe a polar corrugation as the follow
\begin{eqnarray} \label{corrugation2}
r'= r + A \sin(\nu \theta),
\end{eqnarray}
where $\nu$ is the frequency of the corrugation and it must be a positive integer. Casimir energy corresponding to this configuration is
\begin{eqnarray} \label{E-corrugated2}
E_{\mathrm{sp}}^{\mathrm{cor}} =\frac{2}{\pi}\int_{0}^{\pi / 2} E_{\mathrm{sp}}(\theta)  d\theta 
= -\frac{\pi^2}{360 a^2} \big[\frac{\pi r}{2} + \frac{2 A \sin^2 (\pi \nu /4)}{\nu}\big].
\end{eqnarray}
Investigating correction due to finite conductivity with a similar argument leads in the following result
\begin{eqnarray} \label{polar}
 \hspace{-1 cm} \Delta E_{\mathrm{sp}}^{\mathrm{cor}} \thicksim -\frac{\pi^{2}}{60} \big[\frac{\pi r}{2} + \frac{2 A \sin^2 (\pi \nu /4)}{\nu}\big]\sum_{n=1}^{4}\,C_n \,(\frac{\lambda_p}{2 \pi})^{n}\frac{1}{(n+2) a^ {n+2}}.
\end{eqnarray}
\par
\subsection{Effect of a Golf Ball Like Corrugation}
As the final example, we perform the calculations for one more complete case:
\par
A sphere with both azimuthal and polar corrugations. This sinusoidally corrugation is in some way similar to a golf ball and one may describe this corrugation by the following ansatz
\begin{eqnarray} \label{corrugation3}
r'= r + A \sin(\nu \theta) \sin(\mu \varphi).
\end{eqnarray}
The shortest distance is still $a$ even in this case, it is measured from the pole of the sphere (where $\theta,\varphi=0$) to the plate. The Casimir energy associated with this configuration and correction due to finite conductivity can be written as
\begin{eqnarray} \label{E-corrugated3}
E_{\mathrm{sp}}^{\mathrm{cor}}= -\frac{\pi}{720 a^2} \big[\pi ^2 r +\frac{ 4 A \sin ^2(\pi \mu) \sin^2(\pi \nu /4)}{ \mu  \nu} \big],
\end{eqnarray}
considering that for integer $\mu$, $\sin(\pi \mu)=0$ this equation takes the form of Eq. (\ref{E-sp}) and this kind of corrugations dose not contribute either in the Casimir energy of sphere-plate geometry or in the the correction related to finite conductivity. Therefor the finite conductivity correction is the one mentioned in Eq. (\ref{conductivity}) yet.
\par
\section{The Finite Temperature Corrections of the Casimir Energy}
The finite temperature Casimir force per unit area for two parallel plates reads (see \cite{Teo2} and references there in)
\begin{eqnarray} \label{T1}
 \mathfrak{F}_{\mathrm{pp}}(H) = -\frac{\pi ^2}{240 H^4}-\frac{\pi ^2 T^4}{45}+\frac{\pi T}{ H^3} \sum_{k=1}^\infty  \sum_{l=1}^\infty \frac{k^2}{l} \exp(-\frac{\pi k l}{H T}) ,
\end{eqnarray}
or
\begin{eqnarray} \label{T2}
\mathfrak{F}_{\mathrm{pp}}(H) = -\frac{\xi_R(3) T}{4 \pi  H^3}-\frac{2 T}{\pi}\sum_{k=1}^\infty  \sum_{l=1}^\infty \bigg(\frac{2 \pi^2 l^2 T^2}{k H}+\frac{\pi l T}{k^2 H^2}+\frac{1}{4 k^3 H^3}\bigg)e^{-4 \pi H T k l}
\end{eqnarray}
with $H$ as the separation distance between the plates.
\\
\subsection{Low Temperature Limit}
Eq.(\ref{T1}) explains that in the low temperature region thermal correction is dominated by the zero temperature term and therefore the leading term of the thermal corrections in the low temperature limit is
\begin{eqnarray} \label{correction}
\Delta \mathfrak{F}_{\mathrm{sp}}^{\mathrm{T}}\sim -\frac{\pi ^2 T^4}{45}.
\end{eqnarray}
Imposing PFA on this term results in the following thermal correction of the normal Casimir force
\begin{eqnarray} \label{correction of normal f}
\Delta F_{\mathrm{sp}}^{\mathrm{T}}\sim -\frac{2 \pi ^3 T^4 r^2}{45}+ \ldots ,
\end{eqnarray}
which is independent of $a$. Consequently thermal correction of the Casimir energy can be obtained. Taking azimuthal corrugation into account, one obtains the following expression for thermal correction of the Casimir energy as
\begin{eqnarray} \label{correction-phi}
\Delta E_{\mathrm{sp}}^{\mathrm{cor, T}}& \sim &
\frac{2 \pi ^3 T^4 a}{45} \frac{1}{2 \pi} \int_0^{2 \pi} (r+A \sin (\mu \varphi))^2 d\varphi+ \ldots\\ \nonumber
& = & \frac{\pi ^3 T^4 a}{45}\: (A^2+2 r^2)+ \ldots.
\end{eqnarray}
For the investigated polar corrugation we obtain
\begin{eqnarray} \label{correction-theta}
\Delta E_{\mathrm{sp}}^{\mathrm{cor, T}}\sim \frac{\pi ^2 T^4 a}{45}
 \bigg[\frac{ 8Ar \big(1-\cos(\pi \nu /2)\big)+A^2 \pi \nu +2 \pi r^2 \nu}{\nu}\bigg]+ \ldots
\end{eqnarray}
 and for the golf ball like corrugation thermal correction to the Casimir energy in the low temperature region can be obtain in a similar manner.

\subsection{High Temperature Limit}
According to  Eq. (\ref{T2}), in the high temperature region the exponential term goes to zero quickly and therefore the classical term is the first term of the thermal correction
\begin{eqnarray} \label{High correction}
\Delta\mathfrak{F}_{\mathrm{pp}}^{\mathrm{High T}}(H)\sim -\frac{\xi_R(3) T}{4 \pi  H^3} + \ldots .
\end{eqnarray}
Imposing the proximity force approximation to Eq. (\ref{High correction}) with respect to $r>>a$ results in
\begin{eqnarray} \label{high T f}
 \Delta F_{\mathrm{sp}}^{\mathrm{High T}}\sim -\frac{ \xi_R(3) T r}{4 a^2}+ \ldots ,
\end{eqnarray}
performing integral over this correction one can obtain correction to the energy easily.
% this term may be the main term of the normal Casimir force.
Thermal correction of the Casimir energy in the case of azimuthal corrugation at high temperature is
\begin{eqnarray} \label{high T phi}
\Delta E_{\mathrm{sp}}^{\mathrm{High T}}\sim -\frac{ \xi_R(3) T}{4 a} \big(r+\frac{A \sin^2(\mu \pi)}{\mu \pi}) + \ldots ,
\end{eqnarray}
considering that $\mu$ is integer, this kind of corrugations dose not contribute in the high temperature correction of the Casimir energy.
\\
In the case of polar corrugation one may obtain
\begin{eqnarray} \label{high T theta}
\Delta E_{\mathrm{sp}}^{\mathrm{High T}}\sim -\frac{ \xi_R(3) T}{2 a} \big(\frac{r}{2}+\frac{2 A \sin^2(\nu \pi /4)}{\nu \pi})+ \ldots .
\end{eqnarray}
And finally for the the case of golf ball like corrugation, correction to the Casimir energy is as follow
\begin{eqnarray} \label{high T golf}
\Delta E_{\mathrm{sp}}^{\mathrm{High T}}\sim -\frac{ \xi_R(3) T}{4 a} \bigg( r+\frac{4 A \sin^2(\mu \pi) \sin^2(\nu \pi /4)}{\mu \nu \pi} \bigg) + \ldots,
\end{eqnarray}
considering that $\mu$ is an integer,  Eq. (\ref{high T golf}) indicates that this kind of corrugation dose not contribute in the high temperature correction of the Casimir energy.

\section{Conclusion}
This paper is devoted to obtain the Casimir interaction energy between a sinusoidally corrugated sphere and a plate. First, we present a derivation for the Casimir force achieved by the proximity force approximation. Applying this procedure helps us to emit difficulties appear in the scattering matrix method, whereas the proximity force approximation dose not lead in the exact result. This approximation is applicable for $a/r \leqslant 0.00755$ and an experimental accuracy goal of $1\verb"%"$ for a sphere with radius $r$ in the surface-to-surface closest distance $a$ approach from a flat plate. For such short distances the effect of finite conductivity would be non negligible. With respect to the finite conductivity we obtain the corresponding corrections in the Casimir interaction energy. This energy has a strongly dependence on the geometries, number of spatial dimensions, fields and boundary conditions any kind of deformation may conclude in a modification in this interaction energy. In an attempt along these lines, we suppose the sphere to be corrugated and with the aid of the additive summation we investigate the effect of deformation. We consider that, $r a \gg \lambda^2$ and the amplitude of the corrugation is smaller than the other length scales, for validity of PFA. We have also considered thermal corrections of the Casimir force, To reduce the inaccuracy of the result obtained by PFA.

\section{Acknowledgments}
We are grateful to thank  A. Moradian for helpful discussions and correspondence. We would like also to thanks H. Gies,  B. D$\ddot{o}$brich, and  M. Kr$\ddot{u}$ger, for bringing our attentions to the papers \cite{0601094}, \cite{0810.3480}, \cite{1303.2499} respectively.

\nocite{gs}
\nocite{williams}
\nocite{royden}

\bibliographystyle{plain}
\bibliography{martingales}

\end{document}